\documentclass[useAMS,usenatbib,referee]{mn2e}
\usepackage{graphicx}
\newcommand{\oversim}[2]{\protect{\mbox{\lower0.5ex\vbox{%
  \baselineskip=0pt\lineskip=0.2ex
  \ialign{$\mathsurround=0pt #1\hfil##\hfil$\crcr#2\crcr\sim\crcr}}}}}
\newcommand{\simgreat}{\mbox{$\,\mathrel{\mathpalette\oversim>}\,$}} 
\newcommand{\simless} {\mbox{$\,\mathrel{\mathpalette\oversim<}\,$}} 
\title[Low-Mass Star Cluster Formation Efficiency]
{On the infant weight loss of low- to intermediate-mass star clusters}
   \author[C.~Weidner, P.~Kroupa, D.~N\"urnberger and
     M.~Sterzik]{C.~Weidner$^{1}$ \thanks{E-mail: 
   cweidner@astro.puc.cl}, P.~Kroupa$^{2}$ \thanks{E-mail:
   pavel@astro.uni-bonn.de}, D.~E.~A.~N\"urnberger$^{3}$ \thanks{E-mail:
   dnuernbe@eso.org} and M.~F.~Sterzik$^{3}$
   \thanks{E-mail: msterzik@eso.org}\\
$^{1}$ Departemento de Astronom{\'i}a y Astrof{\'i}sica, Pontificia
      Universidad Cat{\'o}lica de Chile, Av. Vicu{\~n}a MacKenna 4860,\\
      Macul, Santiago, Chile\\
$^{2}$ Argelander-Institut f\"ur Astronomie (Sternwarte), Universit\"at
Bonn, Auf dem H\"ugel 71, D-53121 Bonn, Germany\\
$^{3}$ European Southern Observatory, Alonso de Cordova 3107, Santiago, Chile}
\begin{document}
\bibliographystyle{mn2e}
\date{Accepted ??. Received ??; in original form ??}

\pagerange{\pageref{firstpage}--\pageref{lastpage}} \pubyear{2006}

\maketitle

\label{firstpage}
\begin{abstract}
Star clusters are born in a highly compact configuration, typically
with radii of less than about 1 pc roughly independently of
mass. Since the star-formation efficiency is less than 50 per cent by
observation and because the residual gas is removed from the embedded
cluster, the cluster must expand. In the process of doing so it only
retains a fraction $f_{\rm st}$ of its stars. To date there are no
observational constrains for $f_{\rm st}$, although $N$body
calculations by \citet{KAH01} suggest it to be about 20-30 per cent
for Orion-type clusters. Here we use the data compiled by
\citet{TPP97,TPN98,TPN99} for clusters around young Ae/Be stars and by
\citet{DTP04,DTP05} around young O stars and the study of
\citet{DHD99} of OB associations and combine these measurements
with the expected number of stars in clusters with primary Ae/Be and O
stars, respectively, using the empirical correlation between
maximal-stellar-mass and star-cluster mass of \citet{WK05b}. We find
that $f_{\rm st}\,<\,50$ per cent with a decrease to higher cluster
masses/more-massive primaries. The interpretation would be that
cluster formation is very disruptive. It appears that clusters with a
birth stellar mass in the range 10 to $10^{3}\,M_\odot$ keep at most
50 per cent of their stars.
\end{abstract}


\section{Introduction}
The birth sites of star clusters are dense molecular clouds. As the star
formation efficiency, ${\rm SFE} = \epsilon = M_{\rm ecl}/(M_{\rm ecl}
+ M_{\rm gas})$, where $M_{\rm ecl}$ and $M_{\rm gas}$ is
the mass in stars and gas, respectively, just before star formation
ceases, is generally believed to be low, the expulsion of the
surrounding gas leads to the release of a large fraction of the stars
from the clusters. 

The most efficient mechanisms for gas expulsion are the radiation
and winds of massive stars and supernovae. \citet{Kr04b} gives a
simple example which shows that the luminosity output of the OB stars
should be strong enough to destroy their natal cloud, as e.g.~a 15
$M_{\odot}$ star releases as much as $3 \times 10^{50}$ erg per 0.1
Myr into its surrounding medium while a cluster of $10^{4}\,M_{\odot}$
has only a binding energy of $8.6 \times 10^{48}$ erg. Therefore, the
cloud should be dispersed, even before the occurrence of the first
supernova. An observational example is the Orion Nebula Cluster which
is practically gas-free in its centre despite its young age of only
about 1 Myr \citep{HH98}. But strong ``luminosity leakages'' through
low-density holes in the molecular clouds may dampen the effect of
radiation arising from OB stars \citep{DBC05}. In this case,
supernovae may play significant a role in dissolving molecular clouds
\citep{WB77,Go97}.

Rapid removal of gas in the case of $\epsilon < 0.5$ probably leads to
the total destruction of the clusters
\citep{Tu78,Hi80,Ma83,LMD84}. Observations show that
$\epsilon\,\approx\,0.2\,-\,0.4$ \citep{NBYZ02,LL03} but bound star 
clusters like the Pleiades and the Hyades do exist. The problem of
cluster survival after gas expulsion has been identified as one of the
key problems in astrophysical research \citep{DAB06}.

With the use of the \citet{SJA99b} $N$body6 algorithm and including
100 per cent primordial binaries, stellar evolution and a realistic
Galactic tidal field, \citet{KAH01} re-examined the problem of cluster
survival after gas expulsion. They 
showed that from an embedded cluster with $10^{4}$ stars and brown
dwarfs, a bound object with about 25 per cent of the initial number of
stars can survive, even with $\epsilon$ = 0.33 and explosive
residual-gas removal. This remnant cluster is the core of an
  expanding OB association. Thus, the Pleiades and Hyades may have
formed from a compact Orion-Nebula-type object but only retained $f_{\rm
  st}\,\approx\,0.25$ of their birth-stellar mass after removal of 
their residual gas. This fraction, however, depends on the radial
density profile of the pre-expulsion cluster \citep{BK03a,BK03b}, and
also on the relative distribution of the stars and gas \citep{Ad00}.

\citet{KB02} studied the effect of cluster dissolution on the observed
initial cluster mass function (ICMF). They concluded that three
different regimes of clusters may exist which they call type I, II and
III clusters. 
\begin{itemize}
\item[(i)] Type I. Sparse low-mass clusters which contain no O
  stars. They have stellar masses, $M_{\rm ecl}$, below
  $300\,M_{\odot}$ or $N_{\rm ecl}\,<$ 1000 stars. As the stellar
  winds and ionising radiation of the stars is low because of the lack
  of O stars, the gas expulsion time-scale is of the same order as the
  crossing time of the cluster (a few Myr). Thus, \citet{KB02}
  expected $f_{\rm st}\,\approx\,0.5$ for type I clusters.
\item[(ii)] Type II. Clusters with $10^{3}\,\simless\,N_{\rm
  ecl}\,\simless\,10^{5}$ or $300\,\simless\,M_{\rm
  ecl}\,\simless\,30000\,M_{\odot}$. They have only a few O stars. But
  due to their still rather low mass, gas expulsion is 'explosive' on
  a time-scale of a few $10^5$ years. Given the destructive
  residual-gas expulsion \citet{KB02} suggested $f_{\rm
  st}\,\approx\,0.1\,-\,0.2$ for type II clusters.
\item[(iii)] Type III. Massive clusters with more than $N_{\rm
  ecl}\,\simgreat\,10^{5}$, $M_{\rm
  ecl}\,\simgreat\,30000\,M_{\odot}$. These clusters can have
  thousands of O stars but due to the high mass the ionised gas is
  expelled adiabatically, with a time-scale longer than the
  crossing-time of the cluster. If this is correct then $f_{\rm
  st}\,\approx\,0.5$ for type III clusters.
\end{itemize}
The problem \citet{KB02} faced was that $f_{\rm st}$ was virtually
unconstrained by observations.

The aim of this contribution is to show how the fraction of retained
stars for type I and low-mass type II clusters may be observationally
constrained in order to improve our understanding of their very early
evolution and possible fate. To achieve this a large set of
observations addressing the issue of clustering around young
intermediate-mass stars available in the literature
\citep{TPP97,TPN98,TPN99,DTP04,DTP05,DHD99} is used.\\

In Section~\ref{sec:proc} we describe the procedure to calculate the
surviving star fraction. This is followed by Section~\ref{sec:results}
where the results of this work are presented before they are 
discussed in Section~\ref{sec:discuss}.

\section{The Procedure}
\label{sec:proc}

\subsection{The stellar initial mass function}
\label{sub:IMF}
The following multi-component power-law IMF is used to estimate the
number of stars expected in a star cluster:

{\small
\begin{equation}
\xi(m) = k \left\{\begin{array}{ll}
\left(\frac{m}{m_{\rm H}} \right)^{-\alpha_{0}}&\hspace{-0.25cm},m_{\rm
  low} \le m < m_{\rm H},\\
\left(\frac{m}{m_{\rm H}} \right)^{-\alpha_{1}}&\hspace{-0.25cm},m_{\rm
  H} \le m < m_{0},\\
\left(\frac{m_{0}}{m_{\rm H}} \right)^{-\alpha_{1}}
  \left(\frac{m}{m_{0}} \right)^{-\alpha_{2}}&\hspace{-0.25cm},m_{0}
  \le m < m_{1},\\ 
\left(\frac{m_{0}}{m_{\rm H}} \right)^{-\alpha_{1}}
    \left(\frac{m_{1}}{m_{0}} \right)^{-\alpha_{2}}
    \left(\frac{m}{m_{1}} \right)^{-\alpha_{3}}&\hspace{-0.25cm},m_{1}
    \le m < m_{\rm max},\\ 
\end{array} \right. 
\label{eq:4pow}
\end{equation}
\noindent with exponents
\begin{equation}
          \begin{array}{l@{\quad\quad,\quad}l}
\alpha_0 = +0.30&0.01 \le m/{M}_\odot < 0.08,\\
\alpha_1 = +1.30&0.08 \le m/{M}_\odot < 0.50,\\
\alpha_2 = +2.35&0.50 \le m/{M}_\odot < 1.00,\\
\alpha_3 = +2.35&1.00 \le m/{M}_\odot.\\
          \end{array}
\label{eq:imf}
\end{equation}}
\noindent where $dN = \xi(m)\,dm$ is the number of stars in the mass
interval $m$ to $m + dm$. The exponents $\alpha_{\rm i}$ represent the
standard or canonical IMF \citep{Kr01,Kr02}. The advantage
of such a multi-part power-law description are the easy integrability
and, more importantly, that {\it different parts of the IMF can be
changed readily without affecting other parts}. Note that this form is
a two-part power-law in the stellar regime, and that brown dwarfs
contribute about 4 per cent by mass only. A log-normal form below 1
$M_{\odot}$ with a power-law extension to high masses was suggested by
\citet{Ch03}. Today the observed IMF is understood to be an invariant
Salpeter/Massey power-law slope \citep{Sal55,Mass03} above $1\,
M_\odot$, being independent of the cluster density and metallicity for
metallicities $Z \simgreat 0.002$
\citep{MH98,SND00,SND02,PaZa01,Mass98,Mass02,Mass03,WGH02,BMK03,PBK04,PAK06}.

The basic assumption underlying our approach is the notion that all stars
in every cluster follow this same universal IMF, which is consistent
with observational evidence \citep{Elme99,Kr01}.

\subsection{The maximum-star-mass vs cluster-mass relation}
\label{sub:mmax-mecl}
In a series of recent publications \citep{KW03,WKL04,WK04,WK05a,WK05b}
the influence of stars forming predominately in star clusters which
later dissolve into the field was studied. During this process support
for the possibility of a maximum mass for stars was found on a statistical
basis \citep[for R136 in the LMC,][]{WK04}, a result later confirmed
by several independent studies \citep{OC05,Fi05,Ko06}. This work
then yielded to a more thorough investigation of massive stars in star
clusters, resulting in the finding of a probably physical (and not
statistical) relation between the mass of the most massive star in a
young ($< 3\,{\rm Myr}$) star cluster and the mass of the harbouring
star cluster \citep{WK05b}. One consequence of this relation
would be an ordered formation of star clusters meaning that low-mass
stars form first and that star-formation ceases with the appearance of
the high-mass stars. This may be a natural outcome of termination of
star formation by feedback, and had been suggested in the study of the
Hyades and Pleiades by \citet{He62} and in a study of NGC 3293 by
\citet{HM82}.

\begin{figure}
\begin{center}
\includegraphics[width=8cm]{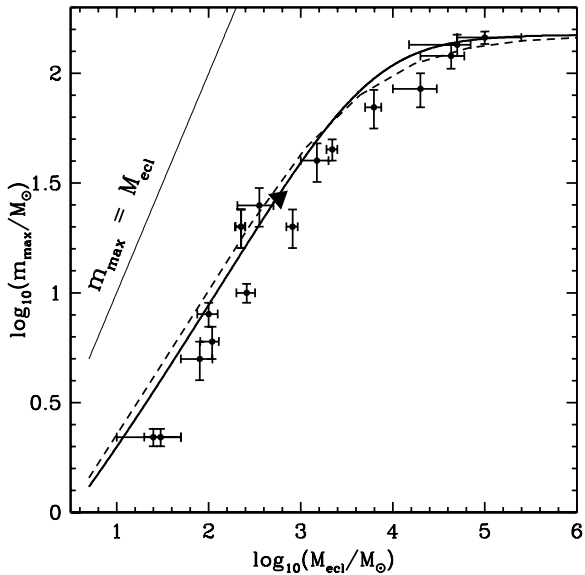}
\vspace*{-2.0cm}
\caption{The logarithm of the most-massive-star versus logarithm of
  the stellar cluster mass. The {\it thick solid line} is the
  semi-analytic result from \citet{WK04}. The {\it thick
  dashed line} is the result of the 'sorted sampling'
  Monte-Carlo experiment of \citet{WK05b}. The {\it dots} with
  error bars are observational values for a sample of young clusters
  from the literature \citep[see][for details on the list]{WK05b}. The
  {\it large triangle} is the most-massive star in a state-of-the-art
  hydrodynamical star cluster formation simulation
  \citep{BBV03,BVB04}. The {\it thin solid line} on the left side of
  the corner, labelled with '${\rm m}_{\rm max} = {\rm M}_{\rm ecl}$',
  indicates the limit where all mass of a star cluster is concentrated
  only in one star.}
\label{fig:mmax-mecl}
\end{center}
\end{figure}
The most-massive-star vs cluster-mass relation ({\it thick solid line}
in Fig.~\ref{fig:mmax-mecl}) then follows by using the cluster mass,
$M_{\rm ecl}$,
\begin{equation}
M_{\rm ecl} = \int_{m_{\rm low}}^{m_{\rm max}} m\,\xi(m)\,dm
\label{eq:Meclmc}
\end{equation}
and taking into account that there exists exactly one most massive
star in each cluster. This condition can be written as
\begin{equation}
1 = \int_{m_{\rm max}}^{m_{\rm max*}} \xi(m)\,dm.
\label{eq:mmmc}
\end{equation}
Here $m_{\rm low}$ = 0.01 $M_{\odot}$ is the minimal fragmentation
mass, $m_{\rm max}$ the most-massive star in a cluster and $m_{\rm
  max*}\,\approx\,150\,M_{\odot}$ the measured maximal stellar mass
limit \citep{WK04,Fi05,OC05,Ko06}. On combining eqs.~\ref{eq:mmmc}
and~\ref{eq:Meclmc} the analytical function 
\begin{equation}
m_{\rm max} = m_{\rm max}^{\rm ana}(M_{\rm ecl})
\label{eq:MmaxvsMecl}
\end{equation}
is quantified by \citet{WK04} and shown as a {\it thick solid line} in
Fig.~\ref{fig:mmax-mecl}.

With this relation we can now calculate the mass of the cluster for an
observed most-massive (primary) star.

\subsection{The data}
\label{sub:data}
\citet{TPP97,TPN98,TPN99} study the clustering of stars around
intermediate-mass pre--main-sequence stars (Herbig Ae/Be stars) with a
large set of near-infrared observations with two different
methods. They looked for over-densities in the photometric $K$-band
flux around these stars in comparison to the background away from the
stars and directly counted the stars found in the $K$-band close to these
stars. Amongst their main results are:
\begin{itemize}
  \item Young star clusters are very compact ($\approx 0.2$ pc).
  \item The more massive a PMS star is the larger is the cluster around it.
  \item There is a gradual change around the spectral type B7 from Ae
  stars without noticeable clusters to Be stars with clusters.
\end{itemize}
Concerning the third result it is interesting to note that the
transition mass is also roughly the same stellar mass above
which standard accretion theory of stars starts to fail. \citet{TPN99}
therefore conclude further that massive star formation ``is influenced by
dynamical interaction[s] in a young cluster environment.``

The data set from \citet{TPN99} already includes completeness limits
down to which the masses of all stars should have been detected by
their method. These limits are calculated for zero and 2 magnitudes of
extinction in the $K$-band. Therefore the observed number of stars
($N_{\rm obs}$ in Tab.~\ref{tab:expec}) would be all stars above the
first limit if no extinction would be present or $N_{\rm obs}$ would
be only the stars above the second limit if there is extinction. But
the data do not allow differentiation between these two cases.

\citet{TPN99} also already converted the colours of the A- and B-stars
into effective temperatures ($T_{\rm eff}$) and luminosities. With the
use of a large set of main-sequence and pre-main-sequence stellar models
\citep{DAM97,HPT00,BM01,MM03} the luminosities and $T_{\rm eff}$
values are converted to stellar masses and ages here. Note that
\citet{TPN98} have determined ages but only for stars below 6
$M_{\odot}$ and with the use of the \citet{PS90,PS93} models. For a
consistent approach all ages are re-derived here with newer
models. Some of the new ages ($\approx\,30\%$ of the stars in the
Testi sample) are substantially larger (more than 200 \%) than the
older ones, but the bulk are reasonably close to the \citet{TPN98}
values. This reflects the still large uncertainties in the ages and
models for PMS stars.

Complementary to the \citet{TPN98} sample, \citet{DTP04,DTP05} searched for
evidence if O stars observed in the field originate from young
star-forming regions or if in-situ field formation of massive stars is
possible. Amongst their 43 candidate stars they found 5 with
previously unknown small clusters surrounding them. These five stars
are included in this study. The given spectral types have been
converted to $T_{\rm eff}$ with the use of table~3 of
\citet{CG91}. The conversion of the luminosities and the $T_{\rm eff}$
into masses and ages are done with the same models as applied on the
\citet{TPN98} sample. Of the remaining 38 stars \citet{DTP04}
could trace back 27 stars to star forming regions - making them
run-away stars which were dynamically ejected from young cluster cores
\citep{RHM01,PAK06}. The nature of the remaining 11 stars is more
puzzling. \citet{DTP04} classify them as O stars formed in isolation,
representing the lower end of the cluster mass function. But they need
a rather shallow cluster mass function with a slope of $\beta $ = 1.7; a
result in contrast to known observational values which are closer to
$\beta$ = 2 for various environments \citep{LL03,HEDM03,ZF99}. Current
star formation theories differ on the question if isolated formation
of massive stars is possible - some argue against it
\citep{BBZ98,SPH00,BB02,BZ05} while others propagate the concept 
\citep{YS02,LKM03}. No unambiguous evidence for such star formation
has been observationally found other than the existence of isolated O
stars.

Using OB associations with known open clusters as cores is even
more difficult than the above described samples. As they are older
($>$ 10 Myr) other dynamical effects (2-body relaxation, few-body
encounters, tidal stripping due to the Galactic tidal field) already
removed additional stars after all gas is lost from the remaining
cluster. Furthermore, the most massive stars may already have exploded
as supernovae - making our method of estimating the cluster mass
through the most-massive star impossible. Also, discrimination between
members of the remaining open cluster and the OB association is
difficult with current data. Additionally, an OB association can be
the result of the dispersion of more than one cluster. This is
actually seen e.g.~in the Monoceros OB1 association which harbours the
open clusters NGC 2264 and Mon R1 and in Perseus OB2 with the clusters
NGC 1333 and IC 348. Nonetheless, using the study of OB
associations by \citet{DHD99}, an attempt is made to compare to the
results obtained from the \citet{TPN98} and the \citet{DTP04} sample.
Particularly suitable from that data set is the Cassiopeia-Taurus
association with the open cluster $\alpha$ Persei at its core. For
Cas-Tau 83 B stars are counted and for $\alpha$ Persei 30 B
stars. This gives a total of 113 B stars of which 27\% ($f_{\rm st}$ =
0.27) are retained in the cluster. An estimate of the original cluster
mass is more difficult. Using an age of 50 Myr \citep{DHD99} for both
cluster and association, the most-massive star still alive should be
around 7.5 $M_{\odot}$ according to the stellar evolution models
described earlier in this work. The border between A and B stars is
around 3.5 $M_{\odot}$. Assuming a canonical IMF (see
Section~\ref{sub:IMF}), 0.68\% of all the stars and brown dwarfs lie
in the mass range between 3.5 and 7.5 $M_{\odot}$. For 113 B
stars this gives a total number of 16600 stars. With a mean stellar
mass of 0.36 $M_{\odot}$ for the canonical IMF, the resulting cluster
mass is about 6000 $M_{\odot}$ in stars. With the same method we
analysed the association -- cluster pairs Cepheus OB2 -- Trumpler 37
and Monoceros OB2 -- NGC 2244 from \citet{DHD99}, supplemented with
data from the WEBDA\footnote{http://www.univie.ac.at/webda//} database
and from \citet{MJD95}. All results are shown in
Table~\ref{tab:expec}.

The relation between the maximum-star-mass and the cluster-mass from
Section~\ref{sub:mmax-mecl} (eq.~\ref{eq:MmaxvsMecl}) is then used to
connect the masses of the most massive stars observed in the combined
\citet{TPN99} and \citet{DTP05} sample with the initial star-cluster
masses. With the use of the canonical IMF (eqs.~\ref{eq:4pow} and
\ref{eq:imf}) the number of stars expected for each cluster, $N_{\rm
  exp}$, is derived for the two mass (completeness) limits, assuming
50\% binaries in both cases. The stars in binaries are assumed to be
chosen randomly from the IMF subject to the mass-constraint imposed by
our relation in Fig.~\ref{fig:mmax-mecl}. Table~\ref{tab:expec} shows
the results of the conversion.

Also shown in Table~\ref{tab:expec} are the radii, $r_{\rm ecl}$, of
the clusters as far as they have been determined by \citet{TPN98} or
\citet{DTP05}. In those cases where no radii were given (marked by
$^{a}$ in Tab.~\ref{tab:expec}) the mean of the other radii has been
chosen. These radii are needed to obtain the two-body relaxation
times, $t_{\rm rel}$, of the cluster with the use of the following
formulae:

\begin{equation}
\label{eq:trel}
t_{\rm rel} = 0.1 \frac{N}{\ln(N)} t_{\rm cr}\,[{\rm Myr}]
\end{equation}
and
\begin{equation}
\label{eq:tcr}
t_{\rm cr} = \frac{2 r_{\rm ecl}}{\sigma_{\rm ecl}} \approx 4
\left(\frac{100 M_{\odot}}{M_{\rm ecl}}\right)^{\frac{1}{2}}
\left(\frac{r_{\rm ecl}}{\rm pc}\right)^{\frac{3}{2}}\,[{\rm Myr}].
\end{equation}

\begin{table}
{\centering \scriptsize
\caption{\label{tab:expec} Number, designation, mass limits with and
  without extinction, $r_{\rm ecl}$ and the observed numbers of stars
  ($N_{\rm obs}$) from \citet{TPN98} and \citet{DTP05}. Age and star
  masses derived from stellar models. $t_{\rm ecl, in}$ and $t_{\rm
  rel, now}$ are calculated from eqs.~\ref{eq:trel} and
  \ref{eq:tcr}. In the first case $N_{\rm exp} 1$ and $r_{\rm
  ecl}\,=\,0.5\,{\rm pc}$ is used, in the second $N_{\rm obs}$ and
  $r_{\rm ecl}$. The initial embedded cluster masses ($M_{\rm ecl}$)
  are from the maximum-star-mass vs cluster-mass relation
  (Section~\ref{sub:mmax-mecl}), while the expected number of stars
  ($N_{\rm exp}$ 1,2) and the transformation factors ($f_{\rm st}$
  1,2) for the \citet{TPN99} and the \citet{DTP05} sample are derived
  as described in Section~\ref{sub:data}.}
\begin{tabular}{ccccccccccccccc}
\hline
Nr.&Des.&limit 1&limit 2&age&mass&$M_{\rm ecl}$&$r_{\rm ecl}$&$t_{\rm
    rel}$&$t_{\rm rel}$&$N_{\rm obs}$&$N_{\rm exp, 1}$&$N_{\rm
  exp, 2}$&$f_{\rm st}$&$f_{\rm st}$\\
&&no ext.&with ext.&&&&obs&in&now&&&&1&2\\
&&[$M_{\odot}$]&[$M_{\odot}$]&[Myr]&[$M_{\odot}$]&[$M_{\odot}$]&[pc]&[Myr]&[Myr]&&&&&\\
\hline
1   & V645 Cyg&    $<$ 0.1$^{b,c}$&      0.5& 2.8&29.2&621.4&0.6&5.2&0.2&   $>$5$^{d}$&588& 136&0.0085&0.037\\
3   & MWC 137&     $<$ 0.1$^{b}$&      0.5& 3.3&17.0&268.5&0.4&4.1&0.9&  $>$59$^{d}$& 265& 61&0.22&0.97\\
5   & BHJ 71&     $<$ 0.1$^{b}$&       0.5& 0.4&12.3&165.4&0.15&3.6&0.05&   4& 168& 39&0.024&0.10\\
7   & AS 310&     $<$ 0.1$^{b}$&       0.5& 4.2&17.5&280.2&0.4&4.2&0.6&  $>$37$^{d}$& 276& 64&0.13&0.58\\
10  & BD+65$^\circ$ 1637&    $<$ 0.1$^{b}$&    0.5& 0.4& 7.5& 78.6&0.4&3.0&1.0&  29& 84&  19&0.35&1.53\\
11  & HD 216629&     $<$ 0.1$^{b}$&    0.5& 0.3&9.2&107.0&0.1&3.3&0.1&  29& 122&  26&0.26&1.12\\
13  & HD 37490&    $<$ 0.1$^{b}$&      0.5& 0.2&10.1&122.7&0.14&3.4&0.1&   9& 127&  29&0.071&0.31\\
18  & XY Per&    $<$ 0.1&        $<$ 0.1& 6.7& 2.9& 18.3&0.08&2.4&0.06&   3&  23&  23&0.13&0.13\\
19  & LkH$\alpha$ 25&   $<$ 0.1&0.16& 0.8& 4.6& 37.4&0.3&2.6&0.5&  11&  43&  28&0.26&0.39\\
22  & LkH$\alpha$ 257&  $<$ 0.1&0.29& 3.2& 2.9& 18.3&0.3$^{a}$&2.4&0.9&  15&  23&  9&0.65&1.67\\
24  & VY Mon&   $<$ 0.1&         $<$ 0.1& 0.4& 5.4& 47.9&0.25&2.8&0.6&  25& 54& 54&0.46&0.46\\
25  & VV Ser&    $<$ 0.1&        $<$ 0.1& 2.2& 3.4& 23.4&0.1&2.5&0.2&  24&  28&  28&0.86&0.86\\
26  & V380 Ori&    $<$ 0.1&      $<$ 0.1& 2.2& 3.2& 21.4&0.3$^{a}$&2.4&0.4&   3&  26&  26&0.12&0.12\\
27  & V1012 Ori&   $<$ 0.1&      $<$ 0.1& 2.0& 3.0& 19.3&0.3$^{a}$&1.6&0.4&   4&  12&  12&0.33&0.33\\
28  & LkH$\alpha$ 218&  $<$ 0.1&0.42& 2.1& 3.5& 24.6&0.3$^{a}$&2.5&0.5&   8&  29&  8&0.28&1.00\\
29  & AB Aur&    $<$ 0.1&        $<$ 0.1& 3.2& 2.7& 16.3&0.3$^{a}$&2.3&0.4&   $>$3$^{d}$&  20&  20&0.15&0.15\\
30  & VX Cas&    $<$ 0.1&       0.11& 2.7& 2.8& 17.3&0.3&2.4&0.8&  13&  21&  18&0.62&0.72\\
31  & HD 245185&    $<$ 0.1&    0.17& 6.9& 2.1& 11.0&0.3$^{a}$&2.3&0.9&  10&  14&  9&0.71&1.11\\
32  & MWC 480&	    $<$ 0.1&     $<$ 0.1& 6.9& 2.1& 11.0&0.3$^{a}$&2.3&0.5&   $>$3$^{d}$&  14&  14&0.21&0.21\\
35  & IP Per&    $<$ 0.1&       0.12& 5.8& 2.2& 11.8&0.3$^{a}$&2.3&0.5&   3&  15&  12&0.20&0.25\\
37  & MWC 758&	    $<$ 0.1&     $<$ 0.1& 5.6& 2.2& 11.8&0.03&2.3&0.02&   $>$2$^{d}$&  15&  15&0.13&0.13\\
39  & HK Ori&   0.11&       0.49& 6.1& 2.1& 11.0&0.3$^{a}$&2.1&0.7&   7&  12&  3&0.58&2.33\\
43  & BF Ori&         $<$ 0.1&  0.14& 4.4& 2.3& 12.7&0.3$^{a}$&2.3&0.5&   4&  16&  12&0.25&0.33\\
\hline
dW1$^{e}$&HD 52266&$<$ 0.1&0.8&2.0&24.0&455.8&0.3$^{a}$&4.3&0.1&11&386&53&0.029&0.21\\
dW2$^{e}$&HD 52533&$<$ 0.1&0.8&1.9&28.8&606.9&0.3&4.7&0.3&36&507&70&0.071&0.51\\
dW3$^{e}$&HD 57682&$<$ 0.1&0.8&2.0&24.0&455.8&0.3$^{a}$&4.3&0.1&11&386&53&0.029&0.21\\
dW4$^{e}$&HD 153426&0.25&0.8&5.0&39.9&1031.0&0.3$^{a}$&3.3&0.1&11&453&117&0.024&0.094\\
dW5$^{e}$&HD 195592&$<$ 0.1&0.8&4.6&39.8&1027.7&0.25&5.5&0.1&26&841&117&0.031&0.22\\
dW1$^{e}$&HD 52266&$<$ 0.1&0.8&2.0&24.0&455.8&0.3$^{a}$&4.3&0.7&113&386&53&0.29&2.13\\
dW2$^{e}$&HD 52533&$<$ 0.1&0.8&1.9&28.8&606.9&0.3&4.7&1.3&284&507&70&0.56&4.06\\
dW3$^{e}$&HD 57682&$<$ 0.1&0.8&2.0&24.0&455.8&0.3$^{a}$&4.3&0.7&113&386&53&0.29&2.13\\
dW4$^{e}$&HD 153426&0.25&0.8&5.0&39.9&1031.0&0.3$^{a}$&3.3&0.5&113&453&117&0.25&0.97\\
dW5$^{e}$&HD 195592&$<$ 0.1&0.8&4.6&39.8&1027.7&0.25&5.5&0.5&164&841&117&0.20&1.40\\
\hline
&$\alpha$ Persei$^{f}$&-&-&50&-&6000&-&-&-&-&-&-&0.26&-\\
&Trumpler 37$^{f}$&-&-&5&-&4500&-&-&-&-&-&-&0.58&-\\
&NGC 2244$^{f}$&-&-&2&-&6200&-&-&-&-&-&-&0.25&-\\
\hline
\end{tabular}
}

$^{a}$ For these clusters no $r_{\rm ecl}$ values were given in
\citet{TPN98} and \citet{DTP05}. Therefore the mean $r_{\rm ecl}$ of
all clusters is used.\\ $^{b}$ For these stars no mass limits were given in
\citet{TPN98}.\\ $^{c}$ In \citet{TPN98} most of the mass limits were
given as $<$ 0.1 $M_{\odot}$. During the calculation of the expected number of
stars these limits were chosen as the hydrogen burning limit of 0.08
$M_{\odot}$.\\ $^{d}$ For these clusters \citet{TPN98} give only a
minimum number of stars observed. Therefore the $f_{\rm st}$ values
are in these cases only lower limits.\\ $^{e}$ The clusters from the
\citet{DTP04,DTP05} sample are included twice as in the paper not
numbers of stars but densities of stars with large error bars are
given. First they are shown with the mean $N_{\rm obs}$ and then with
the maximum $N_{\rm obs}$.\\ $^{f}$ The $f_{\rm st}$ values for these
clusters in OB associations from \citet{DHD99} are determined in a
different way than the rest. See text for details.
\end{table}

In Table~\ref{tab:expec} we quote the estimated initial relaxation
time, assuming the birth-cluster radii are $r=0.5$~pc with $N_{exp, 1}$ 
stars, and the current relaxation time, assuming the measured cluster
radii and the observed number of stars. As is evident, $t_{\rm rel, in}
>$~age in the majority of the cases (66 \%), such that the current low number
of stars cannot be the result of evaporation from the cluster due to
early two-body relaxation. However, $t_{\rm rel, now} <$~age indicates that
the current remnant clusters are relaxation dominated. $N$body
modelling would be required to further quantify the sum of the effect
of loss of stars through gas expulsion and through the later two-body
relaxation after the remnant cluster has re-virialised. But since
cluster disruption through two-body relaxation takes about $20 \times
t_{\rm rel, now}$ in a solar-neighbourhood tidal field, the later
relaxation-driven evaporation would not have had much time to act
significantly. Stars without any evidence for clustering in the
\citet{TPN98} sample have not been included in Tab.~\ref{tab:expec} as
they are likely dynamically ejected from their birth place \citep{PAK06}.

Several of the stars of the \citet{TPN98} sample show considerable
amounts of gas around them. All cases with an amount of gas larger
than 50\% of the initial embedded cluster mass ($M_{\rm ecl}$) have
been excluded from further analysis and are not listed in
Tab.~\ref{tab:expec}. These clusters have not been evacuated from the
gas yet and therefore have not lost stars through this process
yet. Furthermore these large amounts of gas probably hide larger
numbers of stars. In the remaining clusters only small quantities of gas are
still left which may produce extinction but are not important for the
dynamics of the cluster.
\section{Results}
\label{sec:results}
\citet{KB02} used a transformation function, $f_{\rm st, KB}$, to transform
the mass function of embedded clusters (ECMF) into the initial mass function
of bound gas-free star clusters (ICMF). They used the following description:

\begin{equation}
\label{eq:fst}
f_{\rm st, KB}(M_{\rm ecl}) = 0.5 - 0.4\mathcal{G}(lM_{\rm ecl};\sigma_{lM_{\rm
    ecl}};lM_{\rm ecl}^{\rm expl}),
\end{equation}
with
\begin{equation}
\label{eq:gauss}
\mathcal{G}(lM_{\rm ecl};\sigma_{lM_{\rm ecl}};lM_{\rm ecl}^{\rm expl}) =
e^{-\frac{1}{2}\left(\frac{lM_{\rm ecl} - lM_{\rm ecl}^{\rm
      expl}}{\sigma_{lM_{\rm ecl}}}\right)^2},
\end{equation}
and the constants $\sigma_{lM_{\rm ecl}} = 0.5$ and $lM_{\rm ecl}^{\rm
  expl}\,=\,\log_{10}(M_{\rm ecl}^{\rm expl})$ = 4.0. Thus, for
example, at the mass $10^{4}\,M_{\odot}$, $f_{\rm   st,KB}$ = 0.1, ie
a $10^{4}\,M_{\odot}$ cluster only retains 10\% of its stellar
population according to this parametrisation.  

With the values from Table~\ref{tab:expec} it is now possible to define a
similar quantity: but not a transformation function but transformation
factors, $f_{\rm st}$, for each observed star cluster. This is done by
dividing the observed number of stars, $N_{\rm obs}$, from
\citet{TPN99} and \citet{DTP04} by the calculated expected number of
stars, $N_{\rm exp}$. These values are listed in Table~\ref{tab:expec}.

\begin{figure}
\begin{center}
\includegraphics[width=8cm]{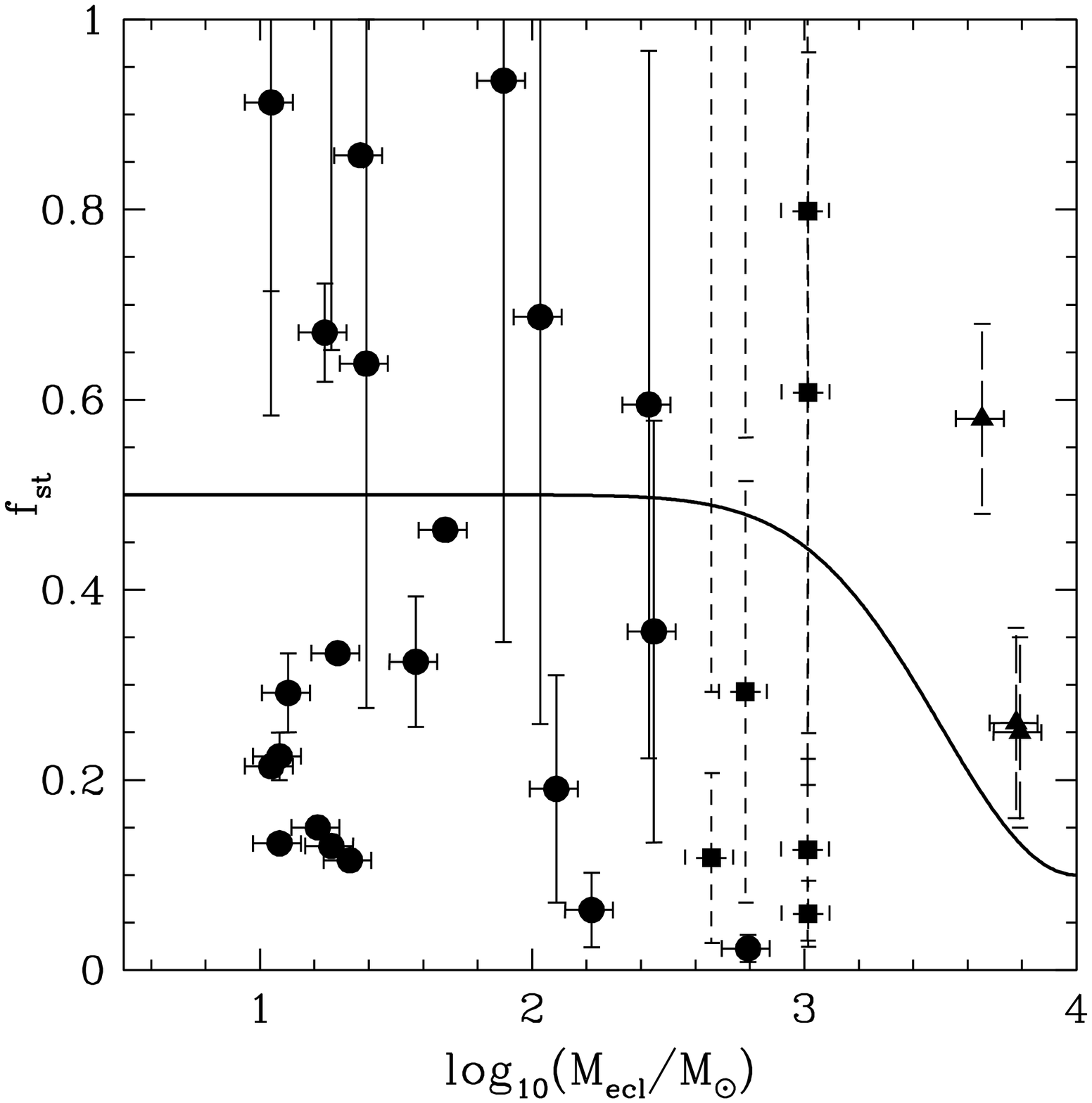}
\vspace*{-2.0cm}
\caption{Transformation factor as a function of embedded cluster mass for the
  sources from the \citet{TPN99} sample ({\it dots} with {\it solid
  error bars}), the \citet{DTP04} sample ({\it boxes} with {\it
  dashed error bars}) and the \citet{DHD99} sample ({\it triangles}
  with {\it long-dashed error bars}). For the \citet{TPN99} and the
  \citet{DTP04} sample the upper ends of the ``error
  bars'' are determined for the cases with assumed extinction, while
  for the lower limit of the ``error bars'' no extinction was
  assumed. The errors for the \citet{DHD99} sample are rather
  arbitrary - simply assuming a square-root (Poisson) error in the observed
  number of OB stars in the association-cluster pairs. The {\it solid
  line} shows the transformation function (eq.~\ref{eq:fst}) adopted
  by \citet{KB02}.}
\label{fig:Meclvsfstg}
\end{center}
\end{figure}

Fig.~\ref{fig:Meclvsfstg} shows the transformation function
(eq.~\ref{eq:fst}) as a {\it solid line} and the derived
transformation factors for each cluster from the \citet{TPN99} sample
as {\it dots} with {\it solid error bars}, from the \citet{DTP04}
sample as {\it boxes} with {\it dashed error bars} and from the
\citet{DHD99} sample as {\it triangles} with {\it long-dashed error
bars}. For the observations the case without extinction is used as
a lower limit of the error bars and the case with extinction as an upper limit.

\begin{figure}
\begin{center}
\includegraphics[width=8cm]{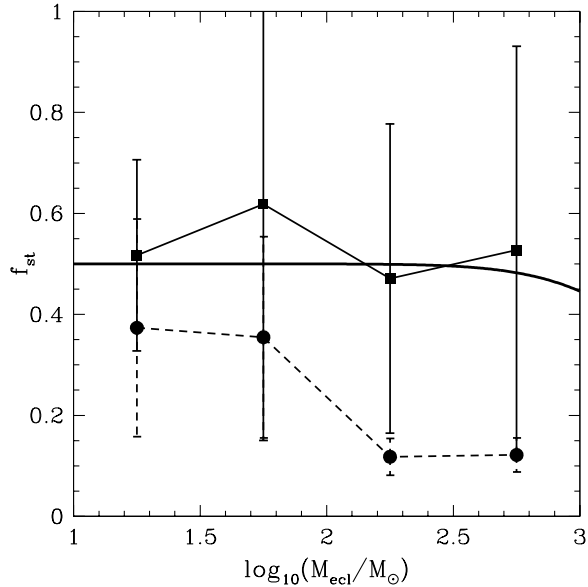}
\vspace*{-2.0cm}
\caption{Like Fig.~\ref{fig:Meclvsfstg} but with \citet{TPN99} data
  binned into four mass bins. The lower limit (no extinction
  corrections, {\it dashed line} connecting the {\it dots}) and the
  upper limit (with extinction corrections, {\it solid line}
  connecting the {\it boxes}) are treated separately. The error bars
  are the variance in the individual mass  bins. The {\it solid line}
  shows the transformation function (eq.~\ref{eq:fst}) adopted by
  \citet{KB02}.}
\label{fig:Meclbin}
\end{center}
\end{figure}

\begin{figure}
\begin{center}
\includegraphics[width=8cm]{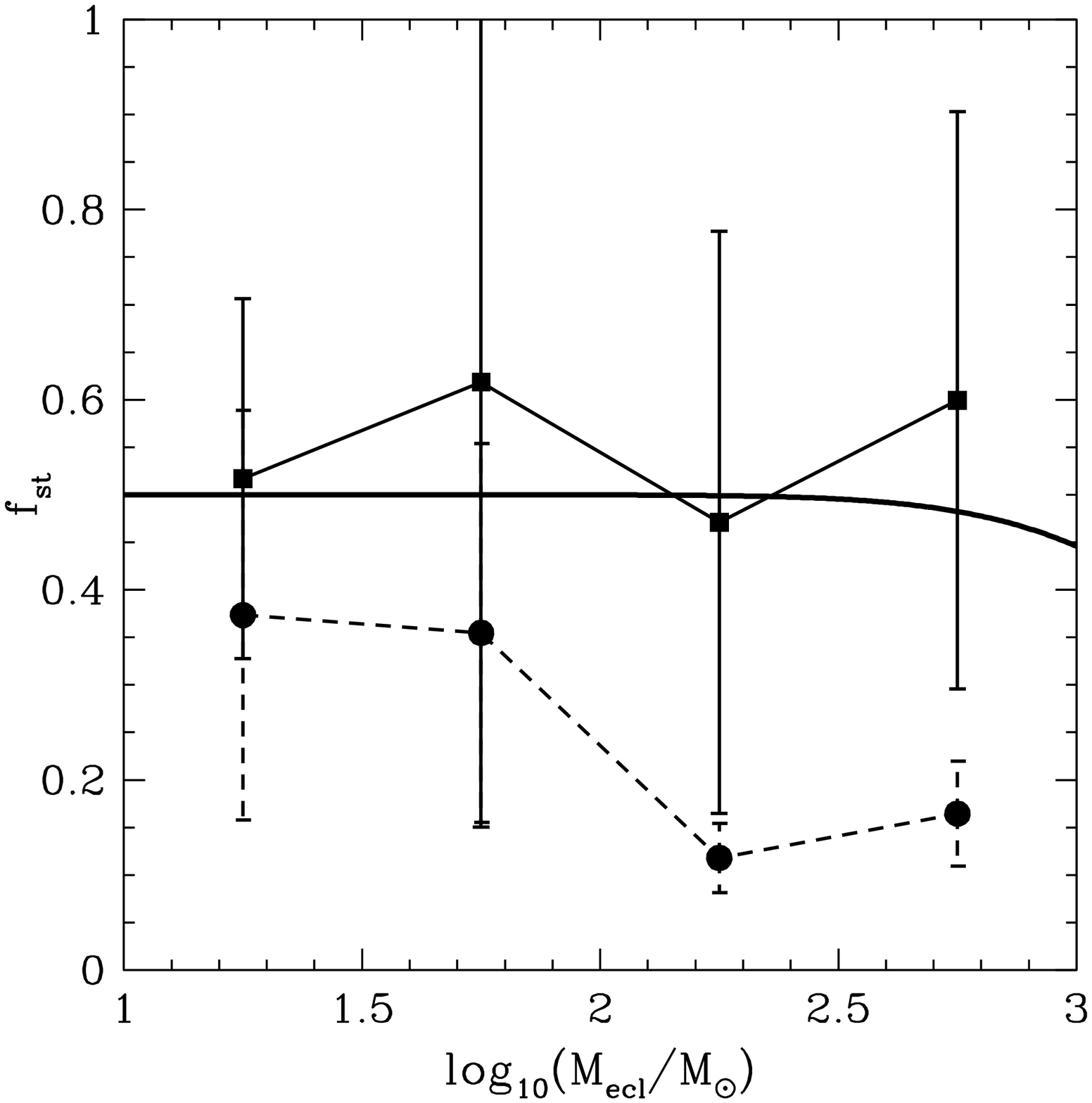}
\vspace*{-2.0cm}
\caption{Same as Fig.~\ref{fig:Meclbin} but with the combined
  \citet{TPN99} and \citet{DTP04} data binned.}
\label{fig:Meclbin2}
\end{center}
\end{figure}

\begin{figure}
\begin{center}
\includegraphics[width=8cm]{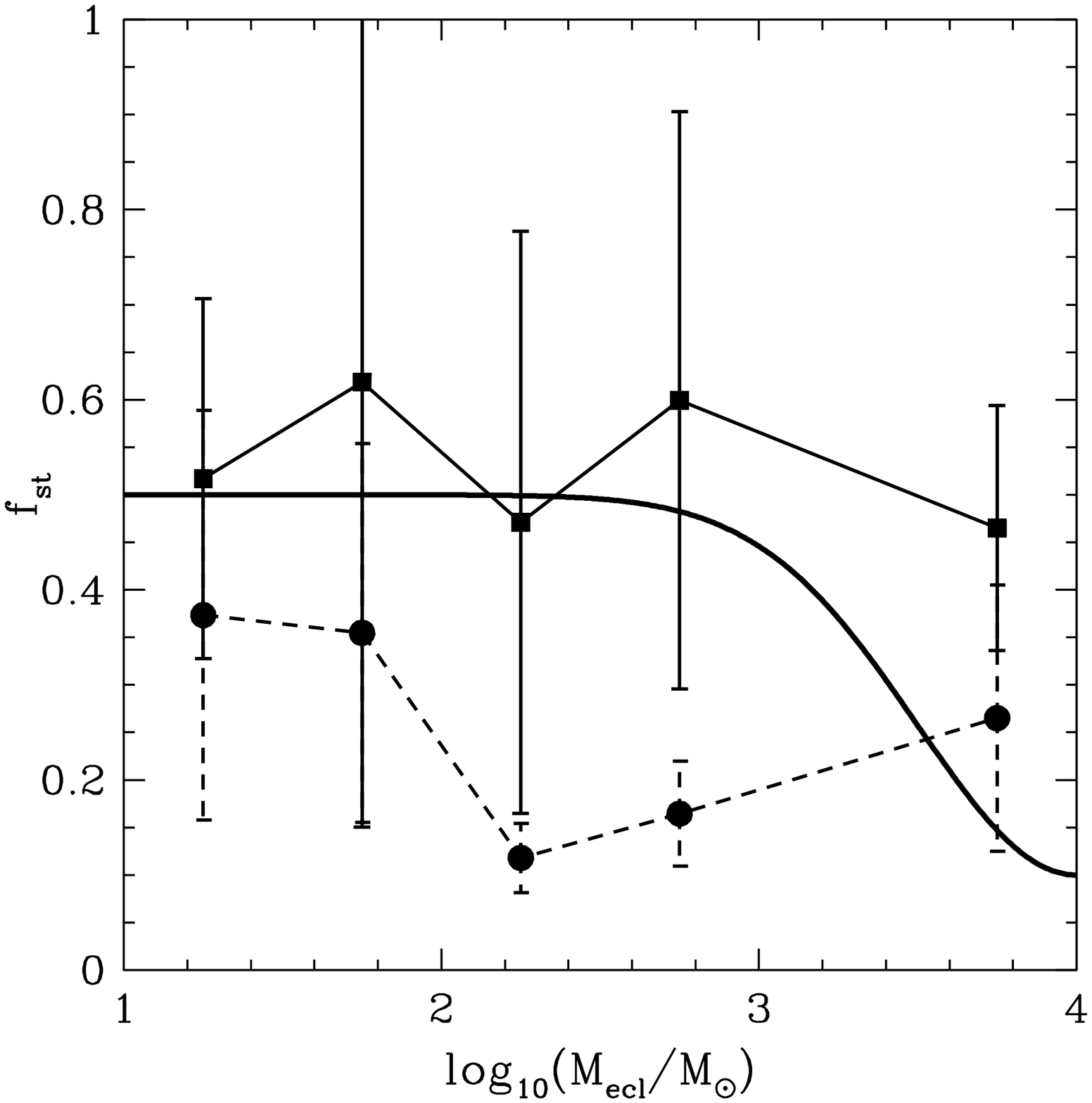}
\vspace*{-2.0cm}
\caption{Same as Fig.~\ref{fig:Meclbin} but with the combined
  \citet{TPN99}, \citet{DTP04} and \citet{DHD99} data binned.}
\label{fig:Meclbin3}
\end{center}
\end{figure}

To reduce the scatter in the observations in Fig.~\ref{fig:Meclvsfstg} the
lower and the upper limits on the $f_{\rm st}$ values are combined
separately into four mass bins. The results are shown in
Fig.~\ref{fig:Meclbin} for the \citet{TPN99} sample only, in
Fig.~\ref{fig:Meclbin2} for the combined \citet{TPN99} and the
\citet{DTP04} sample, and in Fig.~\ref{fig:Meclbin3} for all three
samples combined \citep{TPN99,DTP04,DHD99}. The {\it dots} connected
by a {\it dashed line} are the lower limits and the {\it boxes}
connected by a {\it solid line} are the upper limits. The error bars
for the dots and boxes are the variances in each individual bin. Here
again the {\it solid line} (without dots) is eq.~\ref{eq:fst}, the
transformation function from \citet{KB02}.

The differences between Fig.~\ref{fig:Meclbin} and
Fig.~\ref{fig:Meclbin2} are rather negligible - indicating a good
agreement between the two samples. Because there are substantially
larger uncertainties in the \citet{DHD99} sample (ages, the mass of
the most massive star at 50 Myr, the transition mass between A and B
stars, and the extrapolation from only 0.68 \% of the stars to the
total cluster population), it is not included in the further analysis.

In Fig.~\ref{fig:Meclbin2} it can be seen that the lower limit of the
binned $f_{\rm st}$ factors lie somewhat below eq.~\ref{eq:fst}, while
the upper limits coincide with eq.~\ref{eq:fst}. Therefore it might be
plausible to reduce eq.~\ref{eq:fst} to 0.4 for $M_{\rm
ecl}\,<\,1000\,M_{\odot}$. But the large error bars in this
investigation do not allow a further determination.


In a very recent publication \citep{WL07} the amount of young stellar
objects (YSO) around Herbig Ae/Be stars is further studied with the use of
archival 2MASS and {\it Spitzer} IRAC data. While most of their
targets are already included in this study and the rest is too
obscured by dust, the following additional data for VY Mon and VV Ser
have been extracted. While they observe 26 stars for VY Mon (Testi: 25)
and 22 for VV Ser (Testi: 24), they also give number counts for YSOs
in the whole observed field. In the case of VY Mon they find 42 YSOs
which is quite close to the 54 stars expected to be released by the
cluster through gas expulsion (see Tab.~\ref{tab:expec}). For VV Ser
they find 148 YSOs. This is actually substantially higher than the
expected number of 28, possibly due to other star forming activity
in that region.

\section{Discussion and Conclusions}
\label{sec:discuss}

This contribution shows that with the use of available observational
data \citep{TPP97,TPN98,TPN99,DTP04,DTP05} it is possible to constrain the
star-loss due to gas expulsion in modest star clusters. The
extracted upper and lower limits are shown in
Fig.~\ref{fig:Meclbin2}. The transformation function, $f_{\rm st,
  KB}$, is consistent with the upper end of the data. Thus, clusters
with initial masses in the range 10 - $10^{3}\,M_{\odot}$ appear to
retain about 50\% of their stars, although the uncertainties are
sufficiently large to allow even smaller retainment fractions $f_{\rm
  st}$. Therefore such clusters would appear, in the stages after gas
expulsion, as expanding associations.

In summary, cluster infant weight loss is generally rather high, all
clusters with $M_{\rm ecl}\,\simless\,10^{3}\,M_{\odot}$ losing
$\simgreat$ 50 per cent of their stars (Fig.~\ref{fig:Meclbin2}).
According to \citet{KB02} infant weight loss may increase with cluster
mass, but while being consistent with the present data this is not
required by them, although the dashed constraints in
Fig.~\ref{fig:Meclbin2} may be seen as lending some support to this
notion. Further observations of these and other star-forming regions
are needed to address the fraction of retained stars with more confidence.


\section*{Acknowledgements}
We thank an anonymous referee for valuable comments. This work has
been funded by the Chilean FONDECYT grand 3060096. PK is grateful to
ESO, Santiago, for being supported by a visiting fellowship in
February \& March 2006, when this project was started. This research
has made use of the WEBDA database, operated at the Institute for
Astronomy of the University of Vienna.

\bibliography{WKNSfst.bbl}
\bsp

\label{lastpage}
\end{document}